\documentclass[reprint,prl,superscriptaddress]{revtex4-2}

\usepackage{color}
\usepackage{amsmath,amsthm}
\usepackage{graphicx}

\usepackage{comment}

\newcommand{\av}[1]{\left\langle #1 \right \rangle}
\newcommand{\abs}[1]{\left| #1 \right |}

\newcommand{\up}{\uparrow}
\newcommand{\dn}{\downarrow}
\renewcommand{\Im}{\operatorname{Im}}

\usepackage{tikz}

\usetikzlibrary{calc}
\usetikzlibrary{arrows.meta}

\begin{document}
\title{Larmor precession in strongly correlated itinerant electron systems}
\author{Erik G. C. P. van Loon}
\email{erik.van\_loon@teorfys.lu.se}
\affiliation{NanoLund and Division of Mathematical Physics,
    Department of Physics,
    Lund University, Lund,
    Sweden}

\author{Hugo U. R. Strand}
\email{hugo.strand@oru.se}
\affiliation{School of Science and Technology, Örebro University, SE-701 82 Örebro, Sweden}
\affiliation{Institute for Molecules and Materials, Radboud University, 6525 AJ Nijmegen, the Netherlands}

\begin{abstract}
Many-electron systems undergo a collective Larmor precession in the presence of a magnetic field. In a paramagnetic metal, the resulting spin wave provides insight into the correlation effects generated by the electron-electron interaction. Here, we use dynamical mean-field theory to investigate the collective Larmor precession in the strongly correlated regime, where dynamical correlation effects such as quasiparticle lifetimes and non-quasiparticle states are essential. We study the spin excitation spectrum, which includes a dispersive Larmor mode as well as electron-hole excitations that lead to Stoner damping. We also extract the momentum-resolved damping of slow spin waves. The accurate theoretical description of these phenomena relies on the Ward identity, which guarantees a precise cancellation of self-energy and vertex corrections at long wavelengths. Our findings pave the way towards a better understanding of spin wave damping in correlated materials.
\end{abstract}

\maketitle

This version of the article has been accepted for publication, after peer review but is not the Version of Record and does not reflect post-acceptance improvements, or any corrections. The Version of Record is available online at: \url{http://dx.doi.org/https://doi.org/10.1038/s42005-023-01411-w}

\section{Introduction}

Magnetization dynamics is a field with great technological and scientific importance, with topics ranging from magnetic data storage~\cite{GMR1, GMR2} to magnetic skyrmions~\cite{Fert17}, the ultrafast optical manipulation of magnetism~\cite{Beaurepaire96,Kirilyuk10} and the properties of earth's core~\cite{hausoel2017local}. Although magnetization dynamics in solids is frequently modeled using localized magnetic moments \cite{UppASDBook}, it fundamentally originates from the electrons and their quantum mechanical delocalization and Coulomb interaction.
The description in terms of immobile moments is a particularly drastic approximation for metals with itinerant electrons.

The application of an external magnetic field leads to spin waves in paramagnetic metals. The dispersion relation of the spin wave $\omega_\text{L}(\mathbf{q})$ has a universal value at zero momentum $\mathbf{q}=\mathbf{0}$, corresponding to the Larmor precession of the system's total magnetic moment, while $\omega_\text{L}(\mathbf{q})$ depends on the interactions and correlations in the system at finite $\mathbf{q}$~\cite{Silin58,Silin59,Platzman67,Platzman73,Oshikawa02}. In a Fermi liquid, conduction-electron spin resonance experiments in the presence of a magnetic field can be used to extract the magnetic Landau parameters $B_0$, $B_1$ based on this effect \cite{Platzman67,Schultz67}, which was a crucial confirmation of Fermi liquid theory for the alkali metals in the late 1960's and early 1970's~\cite{Platzman73}. Given the current interest in and limited understanding of non-Fermi liquid behavior in correlated electron systems~\cite{Phillips22,Chowdhury22},
similar studies of spin waves in the presence of a magnetic field have the potential to provide additional insight.

In addition to the dispersion, there is also the question of spin wave damping. For systems with $SU(2)$ spin symmetry, the uniform Larmor precession is undamped, so the damping of the associated spin wave should vanish at long wavelengths ($\mathbf{q} \! \rightarrow \! \mathbf{0}$)~\cite{Platzman73}, similar to plasmons~\cite{GiulianiVignale}. At finite wavelength ($|\mathbf{q}| \! > \! 0$), the collective mode can be damped by electron-hole excitations (Stoner/Landau damping~\cite{GiulianiVignale}) and the finite quasiparticle lifetime of the interacting electrons, i.e., electron-electron scattering.
The latter is a paradigmatic example of a correlation effect, making it hard to model adequately. 

\begin{figure}[h]
\includegraphics{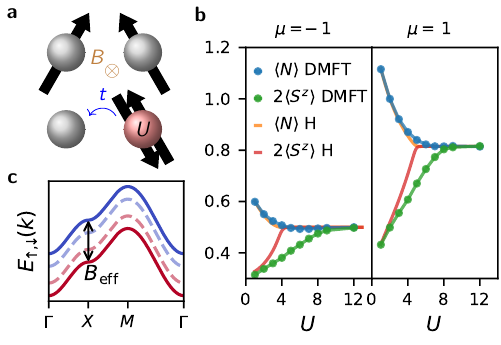}
\caption{
The Hubbard-Zeeman model. \textbf{a} Electrons on a lattice with hopping parameter $t$, Hubbard interaction $U$ and Zeeman magnetic field $B$. \textbf{b}
Density $\av{N} = \av{n_\uparrow} + \av{n_\downarrow} $ in the Hubbard-Zeeman model, calculated using dynamical mean-field theory (DMFT, blue) and the Hartree approximation (H, orange). Similarly, the magnetization $2\av{S^z}=\av{n_\uparrow} - \av{n_\downarrow}$ in DMFT (green) and Hartree (red). Calculations with temperature $T = 1/10$, magnetic field $B=2$ and chemical potential $\mu=\pm 1$. 
\textbf{c} Stoner enhancement of the band splitting. The interaction leads to an enhanced effective Zeeman field $B_\text{eff}$ and therefore a larger splitting between the minority (solid blue) and majority (solid red) bands $E_\up(k)$, $E_\dn(k)$ than in the non-interacting system (dashed blue and dashed red, respectively), where $k$ is taken along the $\Gamma$-X-M-$\Gamma$ high-symmetry path.} 
\label{fig:hubbardzeeman}
\end{figure}

In this work, we study the magnetization dynamics and damping of spin waves in the Hubbard-Zeeman model (Fig. \ref{fig:hubbardzeeman}\textbf{a}). We use Dynamical Mean-Field Theory (DMFT)~\cite{Metzner89,Georges96} to handle strong correlations while maintaining the appropriate Ward identities~\cite{Hafermann14,vanLoon14PRL,Krien17}. 
We show the importance of dynamical correlation effects for the spin excitation spectrum, leading to changes in the dispersion and the damping as well as the appearance of an additional spin wave mode.

\section{Results}

\emph{Larmor precession}
A single magnetic moment in an external magnetic field $B \vec{e}_z$ will partially align itself with the magnetic field, while the in-plane component undergoes Larmor precession with a characteristic frequency $\omega_\text{L} = B$. For a larger system composed of several electrons or magnetic moments, a similar precession occurs for the total magnetization $\vec{S}_T = \sum_a \vec{S}_a$. If the Hamiltonian without field $H_0$ is $SU(2)$ spin-symmetric, and $H = H_0 - B S^z_T$, then the Heisenberg equation of motion for the total magnetization and the precession frequency $\omega_\text{L}=B$ is identical to that of the single moment, independently of the details of the system (see Methods). Here, we have incorporated the $g$-factor into the definition of $B$ and we use $\hbar=1$. The Larmor precession is undamped, a fact that follows directly from the Heisenberg equation of motion.

\emph{Hubbard model} 
We now consider the single-orbital Hubbard model on the square lattice, i.e., $H_0 = -t \sum_{\av{ab},\sigma }c^\dagger_{a\sigma} c^{\phantom{\dagger}}_{b\sigma} + U\sum_{a} n_{a\up}n_{a\dn}$, where $c^\dagger_{a\sigma}$ creates an electron on site $a$ with spin $\sigma \in \{\up,\dn\}$, $c^{\phantom{\dagger}}_{a\sigma}$ is the corresponding annihilation operator, $n_{a\sigma}=c^\dagger_{a\sigma}c^{\phantom{\dagger}}_{a\sigma}$ is a number operator and the sum $\sum_{\av{ab}}$ goes over pairs of neighboring sites $a,b$. The local spin operator $\vec{S}_a$ has components $S_a^{\eta}= \sum_{\sigma\sigma'} c^\dagger_{a\sigma} \sigma^{\eta}_{\sigma\sigma'} c_{a\sigma'}$, where $\sigma^{\eta}$, $\eta \in \{x, y, z\}$ are the Pauli matrices. The parameter $t$ is the hopping matrix element and $U$ is the Coulomb repulsion between electrons on the same site. As $U$ increases, the system becomes more and more correlated. The bandwidth is $8t$ and we use units of energy $t=1$. We consider this model in thermodynamic equilibrium, in the grand canonical ensemble at fixed temperature $T=1/\beta$, chemical potential $\mu$ and magnetic field $B$. The simulations in the figures are for $T=1/10$ and $B=2$. The model is illustrated in Fig.~\ref{fig:hubbardzeeman}\textbf{a}. The magnetization dynamics in equilibrium is encoded in the spatio-temporal correlation function $\chi(t,\mathbf{r}) = \av{S^x(t,\mathbf{r}) S^x(0,\mathbf{0})}$ and its Fourier transform $\chi(\omega,\mathbf{q})$, which we call the susceptibility. We look at the spin components orthogonal to the field since these show the collective Larmor precession. 

Using DMFT, we can calculate both the susceptibility defined above and single-particle properties such as the spectral function $A^\sigma(E,\mathbf{k})$ and the average density and magnetization, $\av{N}$ and $\av{S^z}$, which are shown in Fig.~\ref{fig:hubbardzeeman}\textbf{b}. The DMFT calculations take into account (dynamical) correlation effects such as finite electronic lifetimes, bandwidth renormalization and spectral weight transfer, as well as the associated vertex corrections to the susceptibility. These vertex corrections are needed to satisfy the Ward identities, an exact relation between single-particle and two-particle properties. Only approximations that satisfy the Ward identities are guaranteed to find the undamped Larmor precession at the correct frequency~\cite{Hafermann14,Krien17,KrienThesis} (see Methods). Essentially, the Ward identities lead to a precise cancellation of self-energy effects and vertex corrections for the Larmor precession. We compare the resulting susceptibility to the Random Phase Approximation (RPA). Via the Ward identities, the RPA corresponds to taking into account only static Hartree shifts in the band structure, and it is therefore only applicable to weakly correlated systems.

\begin{figure}
 \includegraphics{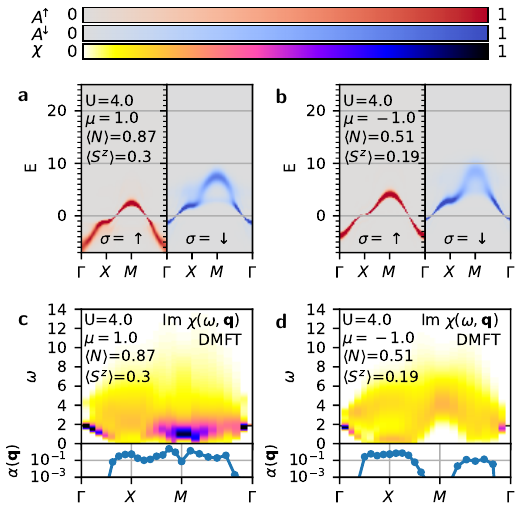}
 \caption{Moderate correlation. \textbf{a,b} Single-particle spectral function $A^\sigma(E,\mathbf{k})$ at a Hubbard interaction $U=4$ for chemical potential \textbf{a} $\mu=1$ and \textbf{b} $\mu=-1$. The associated density $\av{N}$ and magnetization $\av{S^z}$ are indicated. \textbf{c,d} Susceptibility $\Im \chi(\omega,\mathbf{q})$ and damping $\alpha(\mathbf{q})=\partial \Im \chi(\omega,\mathbf{q})/\partial \omega\mid_{\omega=0}$ for \textbf{c} $\mu=1$ and \textbf{d} $\mu=-1$. All momenta are taken along the $\Gamma$-X-M-$\Gamma$ high-symmetry path. Calculations using dynamical mean-field theory (DMFT).}
 \label{fig:susc+spec:4}
\end{figure}
 
\begin{figure}
\includegraphics{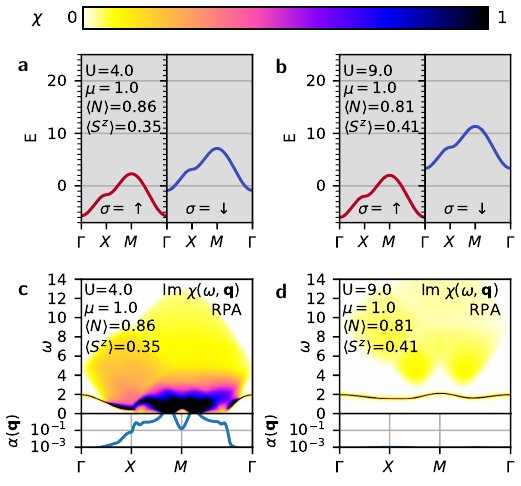}
\caption{Static mean-field. 
\textbf{a,b} Hartree approximation for the band structure at chemical potential $\mu=1$ and a Hubbard interaction \textbf{a} $U=4$ and \textbf{b} $U=9$. Note that the bands are perfectly sharp in the Hartree approximation, so no colorbar is shown. The associated density $\av{N}$ and magnetization $\av{S^z}$ are indicated. \textbf{c,d} Susceptibility $\Im \chi(\omega,\mathbf{q})$ and damping $\alpha(\mathbf{q})=\partial \Im \chi(\omega,\mathbf{q})/\partial \omega\mid_{\omega=0}$ for \textbf{c} $U=4$ and \textbf{d} $U=9$, calculated using the Random Phase Approximation (RPA). All momenta are taken along the $\Gamma$-X-M-$\Gamma$ high-symmetry path. Note that $\alpha(\mathbf{q})<10^{-3}$ for all momenta in \textbf{d}.}
 \label{fig:rpa:susc}
\end{figure}

\emph{Moderate correlation} %
The electron-electron interaction $U$ leads to a finite quasiparticle lifetime for electrons away from the Fermi level, as is visible in the band broadening in Fig.~\ref{fig:susc+spec:4}\textbf{a,b}. Furthermore, there is a Stoner enhancement of the magnetic field (Fig.~\ref{fig:hubbardzeeman}\textbf{c}), which would be given by $B_\text{eff}=B+2U\av{S^z}$ in the Hartree approximation. In fact, this static Hartree mean field overestimates the Stoner enhancement. A minority electron can be on a site precisely at the times when there is no majority electron there. The inclusion of these dynamical correlations reduces the spin polarization $\langle S^z \rangle$, as is visible in Fig.~\ref{fig:hubbardzeeman}\textbf{b}. This is a clear sign of the importance of dynamical correlations even at moderate interaction strengths. Finally, we note the spectral weight transfer away from the main minority spin band, visible in the spectral function $A^\dn(E,\mathbf{k})$ in Fig.~\ref{fig:susc+spec:4}\textbf{a,b}.

Looking at $\chi(\omega,\mathbf{q})$, the spin excitation at the $\Gamma$-point (i.e., $\mathbf{q} = \mathbf{0}$) is fixed at the \emph{bare} Larmor frequency $\omega_\text{L}(\mathbf{0}) = B=2$, as required by the Heisenberg equation of motion and as guaranteed by DMFT's Ward identities, see Fig.~\ref{fig:susc+spec:4}\textbf{c,d}. Universality is lost when moving away from $\Gamma$ to finite momenta $|\mathbf{q}| > 0$. As in the Fermi liquid (see Supplementary Note 1), the Larmor mode disperses downwards, allowing for a determination of Landau parameters~\cite{Platzman67,Schultz67} in principle. At the same time, the electron-hole continuum emanates at $\Gamma$ from the renormalized frequency given by the effective field $B_\text{eff} = B+2U\av{S^z}$. 
Broadening of the Larmor mode due to finite electronic lifetimes is allowed for any $\mathbf{q}\neq \mathbf{0}$, and is clearly visible in Fig.~\ref{fig:susc+spec:4}\textbf{c,d}. Once the Larmor mode and the electron-hole continuum meet further away from $\Gamma$, the Larmor mode can decay into electron-hole pairs (the Stoner/Landau damping), and the mode broadens further. Deep in the Brillouin zone, there is a rich energy and momentum structure in $\chi$ reflecting the particular fermiology at a given filling, including low-energy modes that are indicative of a tendency towards magnetic ordering. Our DMFT simulations have a limited energy resolution and cannot resolve the details of these structures, but they are clearly visible in RPA, see Fig.~\ref{fig:rpa:susc}\textbf{c}. At the same time, RPA does overestimate this tendency towards magnetic ordering~\cite{Vilk97}, which is reflected in the much larger low-energy spectral weight in the RPA susceptibilty. Finally, Fig.~\ref{fig:susc+spec:4}\textbf{c,d} also shows the low-frequency limit of the susceptibility, $\alpha(\mathbf{q})$, which will be discussed in more detail below. The relation to spin-wave spectra in systems with spontaneously broken symmetry~\cite{Niyazi21} is discussed in Supplementary Note 2.

\begin{figure}
 \includegraphics{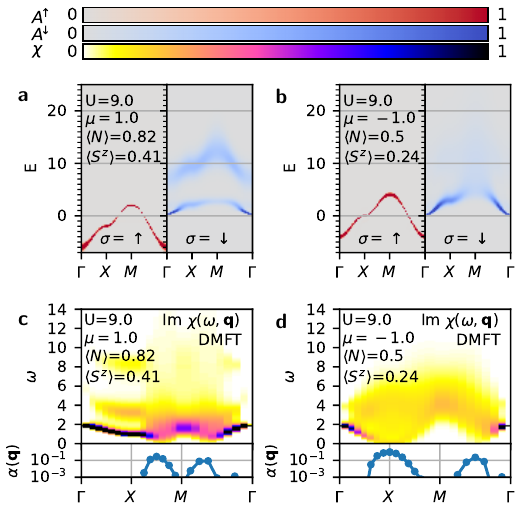}
 \caption{Strong correlation. \textbf{a,b} Single-particle spectral function $A^\sigma(E,\mathbf{k})$ at a Hubbard interaction $U=9$ for \textbf{a} $\mu=1$ and \textbf{b} $\mu=-1$. The associated density $\av{N}$ and magnetization $\av{S^z}$ are indicated. \textbf{c,d} Susceptibility $\Im \chi(\omega,\mathbf{q})$ and damping $\alpha(\mathbf{q})=\partial \Im \chi(\omega,\mathbf{q})/\partial \omega\mid_{\omega=0}$ for \textbf{c} $\mu=1$ and \textbf{d} $\mu=-1$. All momenta are taken along the $\Gamma$-X-M-$\Gamma$ high-symmetry path. Calculations using dynamical mean-field theory (DMFT).}
 \label{fig:susc+spec:9}
\end{figure}

\emph{Strong correlation} %
At larger interaction strengths, the electronic structure changes qualitatively, as shown in Fig.~\ref{fig:susc+spec:9}\textbf{a,b}, becoming reminiscent of a ferromagnetic half-metal~\cite{Katsnelson08}. Due to the large Stoner enhancement of the magnetic field, no minority electrons are present, visible as $\langle N \rangle \approx 2\langle S^z \rangle$ in Fig.~\ref{fig:hubbardzeeman}\textbf{b}. As a result, the majority electrons move as free electrons according to their non-interacting band structure with very long lifetime (no band broadening). For the minority electrons on the other hand, there are both quasiparticle states which are shifted to high energies due to the Stoner enhanced field, and states just above the Fermi level where the minority and majority electrons move coherently so as to avoid doubly-occupying a site. The electron-electron interaction also leads to scattering and thus to a finite lifetime of the minority electrons (band broadening). The precise structure of these features in the minority spectral function depends strongly on the chemical potential $\mu$, which controls the filling of the majority band. A wider set of Hubbard interaction strengths is shown in Supplementary Note 3.

Studying $\chi(\omega,\mathbf{q})$ in Fig.~\ref{fig:susc+spec:9}\textbf{c} for $\mu=1$, two electron-hole continua are visible close to $\Gamma$. One of them emanates from the expected high energy given by the effective magnetic field $B_\text{eff}$, while the one at lower energy originates from transitions between the majority band and the lower minority band. Thus, the spectral weight transfer at the single-particle level is reflected in the spin susceptibility, similar to the observation for plasmons~\cite{vanLoon14PRL}. The appearance of this additional branch is an entirely dynamical correlation driven effect that is beyond Hartree and the corresponding RPA. For $\mu=-1$ in Fig.~\ref{fig:susc+spec:9}\textbf{b,d}, the high-energy branch of the minority single-particle spectrum is very weak, and as a result there are also no high-energy excitations in the susceptibility at $B_\text{eff}\approx 6.3$. 

Although $U=9$ is clearly outside the formal range of applicability of RPA, it is worthwhile to consider how RPA fails by comparing Figs.~\ref{fig:rpa:susc}\textbf{d} and \ref{fig:susc+spec:9}\textbf{c}. The RPA incorrectly predicts an extremely sharp Larmor mode throughout the Brillouin Zone. The reason for this is that RPA only contains damping from electron-hole excitations and it overestimates the energy of the minority electrons substantially, since the Hartree band structure lacks the lower minority branch. 

\begin{figure}
\includegraphics{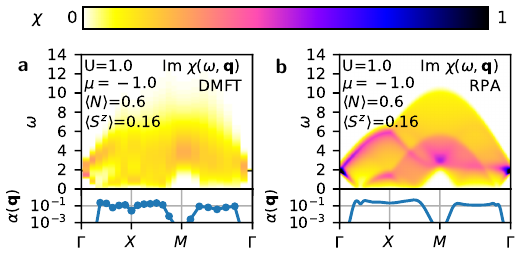}
\caption{Weak correlation. $\Im \chi(\omega,\mathbf{q})$ and damping $\alpha(\mathbf{q})=\partial \Im \chi(\omega,\mathbf{q})/\partial \omega\mid_{\omega=0}$ for \textbf{a} dynamical mean-field theory (DMFT) and \textbf{b} the random-phase approximation (RPA), both at Hubbard interaction $U=1$ and chemical potential $\mu=-1$. The density $\av{N}$, magnetization $\av{S^z}$, susceptibility and damping are similar in both approaches. Note that we are able to perform the RPA calculations with much higher energy resolution. All momenta are taken along the $\Gamma$-X-M-$\Gamma$ high-symmetry path.}
 \label{fig:dmft-rpa:U1}
\end{figure}
  
\emph{Weak correlation} %
For very small values of $U$, RPA is known to be accurate and the comparison of RPA and DMFT can be used to assess the quality of the DMFT calculations. Fig.~\ref{fig:dmft-rpa:U1} shows that both methods indeed give qualitatively similar results: low energy spectral weight in large parts of the Brillouin Zone except close to M, a maximum energy $\omega\approx 6$ at X and one upwards and one downwards dispersing mode close to $\Gamma$. Note that the Stoner enhancement $2U\av{S^z}$ is so small that it is not possible to observe the splitting between the Larmor mode and the electron-hole continuum in either figure. We have a much better energy (and momentum) resolution in RPA, which makes it possible to resolve the sharp features that characterize electronic quasiparticles with a very long lifetime. Note that $U>0$, so DMFT already contains some finite lifetime effects and should not be expected to match the RPA exactly. 

\emph{Damping of slow modes}
Low-energy (slow) magnetization dynamics can be described using the Landau-Lifshitz-Gilbert equation \cite{UppASDBook}, which contains the Gilbert damping as a parameter. In general, the Gilbert damping can be nonlocal, although it frequently assumed to be local ($\mathbf{q}$-independent) for convenience. To connect magnetization dynamics to first-principles calculations, it is necessary to extract the damping parameter from an electronic model. Several mechanisms contribute to the damping of low-energy modes in real materials, including spin-orbit coupling, coupling to the lattice and impurities, decay of spin waves into electron-hole excitations and finite electronic lifetimes. In our model, there is no spin-orbit coupling or coupling to the lattice or impurities by construction, which gives us a clear view on the physics of electron-hole excitations and finite electronic lifetimes. We should stress that spin-orbit coupling is the main damping mechanism for spin waves in many magnetic materials and can in principle be included in multi-orbital DMFT calculations. To connect our electronic model to slow magnetization dynamics~\cite{GilmoreThesis,Umetsu12,Thonig18}, we study $\alpha(\mathbf{q})=\frac{\partial \Im \chi(\omega,\mathbf{q})}{\partial \omega}\mid_{\omega=0}$, which is shown below the color maps in the $\Im \chi(\omega,\mathbf{q})$ figures. Note that we show $\alpha(\mathbf{q})$ on a logarithmic scale with a lower cut-off at $10^{-3}$. 

Starting with the weakly correlated system, Fig.~\ref{fig:dmft-rpa:U1} shows quantitative agreement of $\alpha(\mathbf{q})$ between DMFT and RPA. Since RPA is known to be accurate in this regime, this provides confidence in the numerical methods used to extract $\alpha(\mathbf{q})$ from the analytically continued DMFT spectra. At this small value of $U$, $\alpha(\mathbf{q})$ is relatively flat in a large part of the Brillouin Zone and quickly drops close to $\Gamma$. In RPA, the damping of slow modes $\alpha(\mathbf{q})$ is non-zero when the transferred momenta $\mathbf{q}$ connects the minority and majority Fermi surfaces (see Supplementary Figure 2), since that is where electron-hole damping at $\omega=0$ is possible. For larger values of $U$, $\alpha(\mathbf{q})$ becomes more momentum-dependent, i.e., less local. We find that the RPA does not describe $\alpha(\mathbf{q})$ in moderately and strongly correlated systems accurately. It underestimates the damping by electron-hole excitations since it puts the electron-hole continuum too far way. At strong coupling, it also completely lacks the lower electron-hole excitation branch. As a result, at $U=9$, the RPA predicts a negligible damping of slow modes even though the DMFT calculation shows a substantial damping deep in the Brillouin Zone. At moderate correlation strength, $U=4$, the RPA overestimates the tendency towards magnetic ordering and thereby also $\alpha(\mathbf{q})$.
For additional analysis of the structure of $\alpha$ in real-space see Supplementary Note 4.

\emph{Energy scales}
For clarity, we have illustrated the Larmor precession in situations where the Zeeman splitting is smaller than but comparable to the electronic bandwidth. 
In practice, the Bohr magneton $\mu_B=5\cdot10^{-5}$ eV/T leads to sub-meV Larmor frequencies at realistic magnetic field strengths, substantially below the relevant electronic bandwidth in conventional materials. For that reason, the magnetization dynamics and electronic dynamics can frequently be considered as decoupled phenomena. On the other hand, extremely narrow bands can be created in artificial superlattices, such as twisted bilayer graphene~\cite{Bistritzer11}, transition metal dichalcogenide bilayers~\cite{PhysRevLett.121.026402, Tang:2020aa, Li:2021aa}, and fermions in optical lattices~\cite{Esslinger10}. These systems are promising experimental platforms for studying the collective Larmor precession and the electron-hole continua.

\section{Discussion}

We have studied the collective Larmor precession of correlated electron systems in the presence of a Zeeman field. A comparison of RPA and DMFT calculations shows that dynamical correlation effects are crucial in the moderate and strong correlation regime, for the dispersion and damping of the spin excitations as well as for the single-particle spectrum. The damping is due to a combination of Stoner damping by electron-hole excitations (present but underestimated in RPA), finite electronic lifetimes and vertex corrections (both correlation effects beyond RPA).
The DMFT susceptibility, however, can describe all of these effects, since it satisfies the Ward identities.
The resulting damping of slow modes by electronic excitations is strongly momentum-dependent. 

Current state of the art calculations of the damping of slow modes in the spin dynamics are restricted to static mean-field approximations \cite{UppASDBook, Thonig18}. For this level of theory the torque-torque correlator approach \cite{Kambersky:1976aa} is applicable for damping of spatially uniform spin excitations, i.e.\ at small momenta ($|\mathbf{q}| \approx 0$). However, in several canonical systems like Fe and Ni many-body correlations beyond static mean-field are important \cite{Lichtenstein01,hausoel2017local,Katanin22} and they drastically influence the Gilbert damping \cite{Thonig18}. To account for these effects, ad hoc quasi-particle lifetimes have been introduced in the calculation of the mean-field torque-torque correlator \cite{Kambersky:1976aa, PhysRevLett.99.027204, Thonig18}. Formally this is equivalent to a calculation using the random phase approximation where only the bubble susceptibility is independently modified to account for many-body effects on the single-particle level. However, this completely neglects vertex corrections on the two-particle level, which are a priori just as important, and breaks the Ward identities. 
It also does not take into account the splitting of the minority electron spectral function.
These facts hinder quantitative calculations of the contribution of electronic processes to the Gilbert damping when electronic correlations are important.

For correlated metals like Fe and Ni, the quantitative calculation of Gilbert damping due to electron-electron scattering requires a consistent treatment of single-particle and two-particle correlation effects. In fact, a perfect cancellation of the two effects occurs in the long-wavelength limit, as a consequence of the Ward identities. Hence, we conjecture that DMFT, which satisfies the Ward identity, is able to describe Gilbert damping due to electron-electron scattering accurately. A full account of Gilbert damping needs to take into account spin-orbit coupling as well, since it is the dominant mechanism at long wavelengths and survives in the limit $\mathbf{q}\rightarrow 0$. Spin-orbit coupling can be taken into account in DMFT, so a generalization of the current set-up to multiorbital systems is able to incorporate this mechanism.

\section{Acknowledgments}

The authors would like to acknowledge useful discussions with Danny Thonig, Olle Eriksson, Mikhail Katsnelson and Simon Streib.
EvL acknowledges support from Gyllenstiernska Krapperupsstiftelsen, the Crafoord Foundation and from the Swedish Research Council (Vetenskapsrådet, VR) under grant 2022-03090. EvL also acknowledges support by eSSENCE, a strategic research area for e-Science, grant number eSSENCE@LU 9:1.
HURS acknowledges funding from the European Research Council (ERC) under the European
 Union’s Horizon 2020 research and innovation programme (Grant agreement No.\ 854843-FASTCORR).
The computations were enabled by resources provided by the Swedish National Infrastructure for Computing (SNIC) through the projects
LU 2022/2-32, LU 2021/2-76, 
SNIC 2022/23-304, SNIC 2021/23-370, 
SNIC 2022/21-15, SNIC 2022/13-9, SNIC 2021/28-8,
SNIC 2022/6-113, SNIC 2021/6-133, 
SNIC 2022/1-18, and SNIC 2021/1-36, 
at Lunarc, HPC2N, PDC, and NSC partially funded by the Swedish Research Council through grant agreement no. 2018-05973.

\section{Methods}

\subsection{Dynamical Mean-Field Theory calculations}

\begin{figure}
 \includegraphics{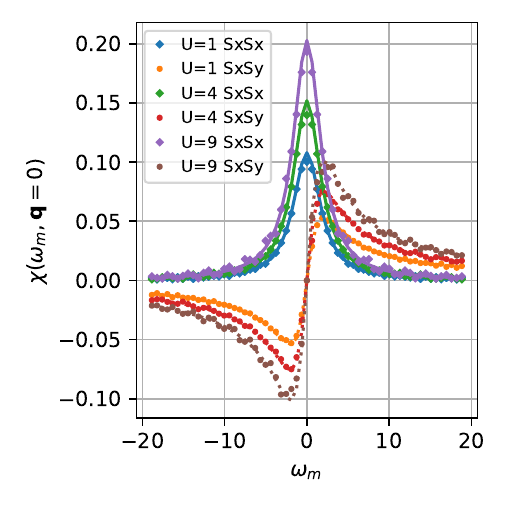}
 \caption{Exact long-wavelength limit. Matsubara axis susceptibility $\chi(\omega_m,\mathbf{q}=0)$ for several values of the Hubbard interaction $U$. The symbols show the calculated dynamical mean-field theory (DMFT) susceptibility, the lines are the exact result of Eq.~\eqref{eq:chisxsx:q0}. Both $\chi^{S^x S^x}$ and $\chi^{S^x S^y}$ are shown.}
 \label{fig:dmft:q0}
\end{figure}

The numerical calculations in this work are based on the Toolbox for Research on Interacting Quantum Systems (TRIQS)~\cite{triqs} and the Two-Particle Response Function toolbox (TPRF)~\cite{Strand:tprf}. The DMFT self-consistency cycle was performed using continuous time hybridization expansion quantum Monte Carlo~\cite{Werner:2006qy, Werner:2006rt, Haule:2007ys, Gull:2011lr} as implemented in TRIQS~\cite{cthyb}, while the two-particle correlation functions of the impurity model were calculated with worm sampling~\cite{PhysRevB.92.155102, PhysRevB.94.125153} using the Wien/Würzburg strong coupling solver (W2Dynamics)~\cite{w2dynamics}. Worm sampling is needed to get access to the full orbital structure of the correlation functions. Based on these ingredients, we calculate the
DMFT lattice susceptibility~\cite{PhysRevLett.69.168, Georges96, PhysRevB.83.085102, Boehnke:2011fk, Boehnke12, PhysRevLett.107.137007, PhysRevLett.109.106401}, 
using an efficient implementation~\cite{technicalpaper} of the Bethe-Salpeter equation to improve the frequency convergence. This implementation corresponds to the dual boson version of the DMFT susceptibility~\cite{Rubtsov12,Hafermann14}, generalized to systems with inequivalent orbitals, and is called the dual Bethe-Salpeter equation (DBSE). This formalism uses the impurity susceptibility $\chi^\text{imp}(\omega_m)$, which in principle can be sampled in the worm CTHYB, but that turned out to be difficult to measure with sufficient accuracy due to ergodicity problems, especially at low frequency. Fortunately, $\chi(\omega_m,\mathbf{0})$ is known exactly (see below). Figure~\ref{fig:dmft:q0} shows the mismatch between the calculated and analytical susceptibility due to the uncertainty in $\chi^\text{imp}$. We introduce a momentum-independent correction, $\chi^\text{corrected}(\omega,\mathbf{q})=\chi(\omega,\mathbf{q}) + \chi^\text{analytical}(\omega,\mathbf{0}) - \chi(\omega,\mathbf{0})$ to enforce the exact result at $\mathbf{q}=0$. This procedure is similar in spirit to the correction applied by Niyazi et al.~\cite{Niyazi21}, although $\chi^\text{analytical}$ is different in this case. 

A momentum discretization of $16\times 16$ is used for the DMFT and DMFT-DBSE calculations, which is sufficient for the chosen parameter regime. The single-particle spectral function $A^\sigma(E,\mathbf{k})$ is plotted on a denser momentum grid for visual purposes, this momentum-grid conversion is possible since $\Sigma$ is local in DMFT.

We note that the calculations presented here take place above $T_\text{c}$, so DMFT's problems with low-temperature spurious ordering~\cite{Schafer15} (at $B=0$) are not an issue. We checked that the leading eigenvalues of the DBSE are below unity to ensure that no ordering phase transition takes place.

The final part of the calculation is the maximum entropy (MaxEnt) analytical continuation~\cite{Jarrell:1996fj} of the susceptibility $\chi^\text{corrected}(\omega_m,\mathbf{q})$ and the single-particle Green's function from Matsubara to real frequency. For this, we use the Python package for analytic continuation (ana\_cont)~\cite{anacont}. 
Analytical continuation from Matsubara to real frequency data is an intrinsically hard computational problem and many methods to perform the continuation have been discussed in the literature, also for susceptibilities (i.e., bosonic functions)~\cite{Gunnarsson10,Schott16,Huang22}. The result of an analytical continuation is in some sense only the most probable result consistent with the Matsubara data. Here, we have performed MaxEnt continuation with an automatic tuning of the hyperparameter, as implemented in ana\_cont. A constant noise amplitude $10^{-3}$ is assumed for the susceptibility ($10^{-2}$ for the spectral function), results were checked to be qualitatively stable under variation of this parameter. A flat default model was assumed.

It is useful to discuss which features in our analytically continued spectra can be relied upon and which properties are more uncertain. Analytical continuation is especially difficult when the Matsubara data is subject to statistical noise, as in our DMFT calculations. Here, we have the advantage that we know the analytical form at $\mathbf{q}=0$ and we have used this to remove the dominant Monte Carlo uncertainty (see above). Thereby, we recover the expected sharp Larmor peak at $\mathbf{q}=0$. We also observe the correct number of spin excitation branches based on the single-particle spectra. The continuation at different momenta is performed independently, and the lack of discontinuities is a good sign for the quality of the continuation. On the other hand, the continuation is likely to blur sharp structures within a region of finite spectral weight, making it hard to identify possible splitting within the collective modes~\cite{McNiven22} that has been discussed for charge fluctuations in the Hubbard model. Analytical continuation is known to be more reliable at low energies than at high energies. This makes the extraction of $\alpha(\mathbf{q})$ relatively safe. On the other hand, in Fig.~\ref{fig:susc+spec:9}\textbf{b} at $\mu=-1$, very diffuse high-energy weight is observed in $A^\dn(E,\mathbf{k})$ and no spectral weight is observed in $\Im \chi(\omega,\mathbf{q})$ in Fig.~\ref{fig:susc+spec:9}\textbf{d}. Given the limitations of analytical continuation, it is hard to rule out that the true $\Im \chi(\omega,\mathbf{q})$ at these parameters has a small amount of high-energy spectral weight. 

\subsection{Random Phase Approximation calculations} 

We have also calculated the susceptibility in the RPA. This takes into account the Hartree renormalization of the band structure and the corresponding static, unrenormalized vertex only, making it inapplicable in the strongly correlated regime. It has the advantage that exact, high-resolution simulations can be done easily, so it is valuable as a reference. DMFT, on the other hand, includes local dynamical correlations, including finite electronic lifetime effects and a local dynamical vertex, so it can describe the strong correlations that we are interested in here. DMFT simulations are computationally more expensive and require analytical continuation (see above), substantially limiting the attainable energy resolution.
We note that both the RPA and DMFT susceptibilities satisfy the Ward identities, which guarantees exact relations at long wavelengths, where Larmor precession takes place. 
Supplementary Note 1 contains a study of the homogeneous electron gas using RPA, which clearly illustrated the physical content of the approximation. 

\subsection{Larmor precession}

General properties of many-body Larmor precession in the presence of a Zeeman field have been studied in several works, see, e.g., Refs.~\cite{Platzman73,Oshikawa02,KrienThesis}. Here, we provide the necessary derivations in a form that closely matches the DMFT results presented in this work.

For a lattice model, Larmor precession in the presence of a Zeeman field occurs for the total magnetization $\vec{S}_T = \sum_a \vec{S}_a$, where $\vec{S}_a$ is the spin on site $a$ and the sum runs over the entire lattice. We consider a Hamiltonian $\hat{H} = \hat{H}_0 - B \hat{S}^z_T$, where $H_0$ is the $SU(2)$ symmetric Hamiltonian without the field, we have incorporated the $g$-factor into the definition of $B$, and $\hbar=1$. The Heisenberg equation of motion for the total magnetization $\langle \vec{S}_T \rangle$ is identical to that of a single magnetic moment
\begin{align}
\frac{d}{dt} \av{\hat{S}^x_T} &= i \av{[\hat{H},\hat{S}^x_T]} = -iB \av{[\hat{S}^z_T,\hat{S}^x_T]}= B \av{\hat{S}^y_T}, \notag \\ 
\frac{d}{dt} \av{\hat{S}^y_T} &=-B\av{\hat{S}^x_T}, \notag \\
\frac{d^2}{dt^2} \av{\hat{S}^x_T} &= - B^2 \av{S^x_T},
\label{eq:sxy_eom}
\end{align}
giving the precession frequency $\omega_\text{L}=B$. Note that the relation $\omega_\text{L}=B$ for the precession of the total magnetization is completely independent of the total number of electrons, the lattice structure and the Coulomb interaction between the electrons. 

In this work, we study many-electron systems on a lattice at finite temperature, i.e., in a statistical ensemble.
For such a statistical ensemble, $\av{S^x}=0$ at any time by symmetry. The Larmor precession is visible in the periodicity of the time-dependent spin correlation function. Since the total spin $S_T$ precesses with the same frequency $\omega_\text{L}$ in all the members of the ensemble, the dynamical total spin-total spin correlation function has a single sharp, undamped mode at the frequency $\omega=\pm \omega_\text{L}=\pm B$. 

Since it relates to the total spin, the Larmor mode is visible at the momentum $\mathbf{q}=\mathbf{0}$ of the dynamical, momentum-dependent susceptibility. We use the general notation $\chi^{AB}(\omega,\mathbf{q})=\av{A B}_{\omega,\mathbf{q}}- \delta_{\omega}\delta_{\mathbf{q}}\av{A}\av{B}$ to denote susceptibilities corresponding to different observables $A$, $B$. We use both real frequencies $\omega$ and Matsubara frequencies $i\omega_m$, for the retarded and Matsubara correlation functions, respectively. In the main text, $A=B=S^x$ and the superscript is dropped for clarity. We also note that $\chi$ is sometimes defined with an overall minus sign e.g., Giuliani and Vignale~\cite{GiulianiVignale} or Krien et al.,~\cite{Krien17}, which requires some care when comparing formulas with the literature. 

The exact form of the dynamical spin susceptibility at $\mathbf{q}=\mathbf{0}$ is~\cite{KrienThesis}
\begin{align}
 \chi^{S^x S^x}(\omega,\mathbf{0}) &= \frac{B \av{S_z}}{B^2-\omega^2}, \label{eq:chisxsx:q0} \\
 \chi^{S^x S^y}(\omega,\mathbf{0}) &= -\frac{i\omega \av{S_z}}{B^2-\omega^2}.
\end{align}
The location of the two poles in the complex $\omega$ plane is known from the equation of motion, and their residue will be derived below using formulas for the large-frequency asymptotics. Both expressions are exact, since they are derived from the equation of motion without approximation. They depend explicitly on $B$ and implicitly on all other system parameters via $\av{S^z}$, which only determines the overall magnitude. Note that there is no discontinuous $\delta_\omega$ contribution~\cite{Hafermann14,Watzenbock22} to $\chi^{S^x S^x}$, since $S^x$ is not a conserved quantity in the presence of a Zeeman field. For $\mathbf{q}\neq 0$, the equation of motion generates additional terms $[H_0,S^x_\mathbf{q}]$, which leads to a dispersion relation $\omega_\text{L}(\mathbf{q})$ that is sensitive to details of the system~\cite{Platzman73}. 

\subsection{Larmor precession: Magnitude of susceptibility}

The frequency structure of the Larmor precession follows directly from the equation of motion. Here, we derive the magnitude of the susceptibility in the Matsubara formalism. The susceptibility is known to decay algebraically at large Matsubara frequencies and the prefactors of this asymptotic decay are given by commutation relations~\cite{Hugel16,Krien17}. For a general susceptibility $\chi^{AB}$, the leading asymptotic Matsubara decay (first moment) is determined  by the commutator $[A,B]$, i.e., $i\omega_m \chi^{AB}(i\omega_m) \rightarrow \av{\left[ A,B\right]}$ for $|\omega_m| \rightarrow \infty$. The relevant case for us is $[S^x,S^y]=iS^z$,
\begin{align}
\chi^{S^x S^y}(i\omega_m,\mathbf{0}) \overset{|\omega_m|\rightarrow\infty}{=}& \av{S^z}/\omega_m.
\end{align}
For $\chi^{S^x S^x}$ and $\chi^{S^z S^z}$, where $A=B$, this commutator vanishes and the next term in the asymptotic expansion is $(i\omega_m)^2 \chi^{AB}(i\omega_m) \rightarrow -\av{\left[[ A,H],B]\right]}$.
Since $H$ appears linearly, the high frequency tail coefficient can be decomposed into contributions from the interaction, the dispersion and the Zeeman field. In the present case, the interaction does not contribute, while the dispersion leads to a coefficient related to $\partial^2 \epsilon_k/\partial k^2$ and is the same for $\chi^{S^x S^x}$ and $\chi^{S^z S^z}$. This term is $\mathbf{q}$-dependent with a vanishing value at $\mathbf{q}=0$. Finally, the Zeeman field is not relevant for $\chi^{S^z S^z}$ but is relevant for $\chi^{S^x S^x}$, where it creates a local term (independent of $\mathbf{q}$). Since this is the only contribution at $\mathbf{q}=0$, using $[[S^x,-B S^z],S^x]=iB[S^y,S^x] = B S^z$, we get
\begin{align}
 (i\omega_m)^2 \chi^{S^x S^x}(i\omega_m,\mathbf{0}) &\overset{|\omega_m| \rightarrow \infty}= -B \av{S^z}.
\end{align}
Higher-order moments of the susceptibility are obtained by inserting additional commutators with the Hamiltonian, i.e., \begin{align}
[[[S^x,H],H],S^x]&=B^2 [[[S^x,S^z],S^z],S^x]=0 \\                                                                                                              
[[[[S^x,H],H],H],S^x]&=-B^3 [[[[S^x,S^z],S^z],S^z],S^x] \notag \\
&= -B^3 (-i) [[[S^y,S^z],S^z],S^x] \notag \\                                                                                                 
&= -B^3 [[S^x,S^z],S^x] \notag \\                                                                                                 
&= -B^3 (-i) [S^y,S^x] \notag \\   
&= B^3 S^z.
\end{align}
Since the only non-commuting part of the Hamiltonian is $-B S^z$, every additional commutator leads to a switch between $S^x$ and $S^y$ and a factor $+i$ or $-i$. Thus, for $\chi^{S^x S^x}$, only the odd commutators survive and the moment-expansion is
\begin{multline}
  \chi^{S^x S^x}(i\omega_m,\mathbf{0})
  =
  -\frac{B \av{S^z}}{(i\omega_m)^2} \sum_{M=0}^\infty \frac{B^{2M}}{(i\omega_m)^{2M}}
  \\
  = -\frac{B \av{S^z}}{(i\omega_m)^2-B^2}.
\end{multline}
By analytical continuation $i\omega_m\rightarrow \omega$, this completes the proof of Eq.~\eqref{eq:chisxsx:q0}. 

Note that the magnitudes of $\chi^{S^x S^x}$ and $\chi^{S^x S^y}$ are linked, since the corresponding observables are governed by the coupled differential equations in Eq.\ (\ref{eq:sxy_eom}). Using $2S^x=S^++S^-$, $2i S^y=S^+-S^-$, Eq.~\eqref{eq:chisxsx:q0} implies that $\chi^{-+}(\omega,\mathbf{0})$ and $\chi^{+-}(\omega,\mathbf{0})$ each have a single simple pole, at $\omega=+\omega_\text{L}=+B$ and $\omega=-\omega_\text{L}=-B$, respectively. These two susceptibilities are related by time-reversal symmetry and $B\mapsto -B$. The commutation relation $[S^+,S^-]=2S^z$ constrains the instantaneous two-particle correlation functions  (that is, the integral over $\omega$ of Eq.~\eqref{eq:chisxsx:q0}) in terms of the one-particle expectation value $\av{S^z}$, and gives a simple explanation for the appearance of $\av{S^z}$ on the right-hand side of Eq.~\eqref{eq:chisxsx:q0}.

\subsection{Larmor precession: limits}

It is useful to verify Eq.~\eqref{eq:chisxsx:q0} in several relevant limits, see also the discussion in Refs.~\cite{KrienThesis,Niyazi21}. 
The limit $\omega\rightarrow 0$ of a dynamic susceptibility is the linear response to an applied field. Here, starting with $B \hat{e}_z$ and applying a small field $\delta B \, \hat{e}_x$ does not change the magnitude of the field to linear order in $\delta B$, but it changes its orientation to $B\hat{e}_z+\delta B \hat{e}_x$, again keeping terms to linear order in $\delta B$ only. The magnetization will follow this orientation, so $\delta\! \av{S^x}=\frac{\delta B}{B} \av{S^z}$. In other words, $\frac{d\av{S^x}}{dB_x}=\frac{\av{S^z}}{B}=\chi^{S^xS^x}(0,\mathbf{0})$, consistent with Eq.~\eqref{eq:chisxsx:q0}. 

If we now take the limit $B\rightarrow 0$, we recover $\chi^{S^x S^x}(0,\mathbf{0})= \frac{d\av{S^z}}{{dB}}=\chi^{S^z S^z}(0,\mathbf{0})$, as expected since the Hamiltonian is $SU(2)$ symmetric in this limit and we are considering a paramagnetic state. Furthermore, the poles in $\chi^{S^z S^z}$ move towards the real axis and at $B=0$, $\chi^{S^z S^z}(\omega,\mathbf{0}) \propto \delta_{\omega}$ as usual for a conserved quantity.

\subsection{Hartree and RPA}

The susceptibility in the non-interacting model is given by the Lindhard bubble, for the spin-flip channel it is 
\begin{align}
  \hat{\chi}_0^{-+}(\omega,\mathbf{q}) = \sum_\mathbf{k} \frac{n^\dn_{\mathbf{k}+\mathbf{q}}-n^\up_{\mathbf{k}}}{\omega-E^\dn_{\mathbf{k}+\mathbf{q}}+E^\up_{\mathbf{k}}}.
\end{align}
At $\mathbf{q}=0$, the denominator simplifies to $E^\dn-E^\up = B$ independent of $\mathbf{k}$, so the momentum sum can be performed to give $\hat{\chi}_0^{-+}(\omega,\mathbf{q}=0)= -2 \av{S^z} / (\omega-B)$. By symmetry, $\chi_0^{+-}(\omega,\mathbf{q}=0)=+2 \av{S^z} / (\omega-B)$, giving the expected result 
\begin{align}
 \hat{\chi}_0^{S^x S^x}(\omega,\mathbf{q}=0) &= \frac{\hat{\chi}_0^{-+} + \hat{\chi}_0^{+-}}{4} = \frac{B \av{S^z}}{B^2-\omega^2},  \\
 \hat{\chi}_0^{S^x S^y}(\omega,\mathbf{q}=0) &= \frac{\hat{\chi}_0^{-+} - \hat{\chi}_0^{+-}}{4i}= \frac{-i\omega \av{S^z}}{B^2-\omega^2}.
\end{align}

The effect of the Hubbard interaction at weak coupling can be studied within the Hartree approximation, which is exact to linear order in the interaction (the Fock self-energy is zero in the Hubbard model). The Hartree self-energy states that the electrons feel an effective additional on-site potential given by the density of the opposite spin, $\Sigma_\sigma = U \av{n_{-{\sigma}}}$. The corresponding enhancement of the band splitting is $\Delta B^\text{H}= 2 U \av{S^z}$, leading to a larger magnetization. This is essentially just the Stoner mechanism~\cite{GiulianiVignale,UppASDBook,Katsnelson08}. Note that this system of equations has to be solved self-consistently. 

The Random Phase Approximation provides us with the dynamical susceptibilities associated with the Hartree single-particle energies. Due to the enhanced band splitting, $\hat{\chi}_0^{-+}$ has its pole at $\omega = B+2U \av{S^z} > \omega_\text{L}$. The RPA susceptibility, written in a tensor formalism, is calculated from the Lindhard bubble via the Bethe-Salpeter equation $\hat{\chi}(\omega,\mathbf{q}) = \hat{\chi}_0(\omega,\mathbf{q}) + \hat{\chi}_0(\omega,\mathbf{q}) \ast \hat{U} \ast \hat{\chi}(\omega,\mathbf{q})$, where $\chi_0$ is the Lindhard bubble. Importantly, the RPA also does not couple different momenta, so the analysis only requires the Lindhard bubble at $\mathbf{q}=0$. 
The $\chi^{-+}$ channel does not couple to other channels and the Hubbard interaction enhances the magnetic susceptibility (Stoner enhancement). 

The solution of the Bethe-Salpeter equation is 
\begin{align}
 \hat{\chi}(\omega,\mathbf{q}) &= \hat{\chi}_0(\omega,\mathbf{q}) \times \left[\hat{1} - \hat{U} \hat{\chi}_0 (\omega,\mathbf{q})\right]^{-1}, \notag \\
 \hat{1} - \hat{U} \hat{\chi}_0 (\omega,\mathbf{q}=0) &\overset{\chi^{-+}}{\rightarrow} 1 - U \frac{-2 \av{S^z}}{\omega - (B+2U\av{S^z})} \notag \\
 &= \frac{\omega - B - 2 U \av{S^z} + 2 U \av{S^z} }{\omega - B - 2 U \av{S^z} } \notag \\
 &= \frac{\omega-B}{\omega - B - 2 U \av{S^z}}, \\
 \chi^{-+}(\omega,\mathbf{q}=0) &= \frac{-2 \av{S^z} }{\omega- B - 2 U \av{S^z} }\times \frac{\omega - B - 2 U \av{S^z}}{\omega-B} \notag \\
 &= \frac{-2 \av{S^z} }{\omega-B} .
\end{align}
The final result for $\hat{\chi}$ indeed has a simple pole at the correct frequency $\omega_\text{L} = B$ and with the appropriate residue required by the equation of motion. The correct result arises from a cancellation between single-particle renormalization and vertex corrections. This cancellation holds more generally, as shown below.

\subsection{Ward identity and Larmor frequency}

In general, the Bethe-Salpeter equation is a matrix equation in fermionic frequency and momentum space. In the RPA, the frequency dependence is trivial: the vertex is frequency-independent and the electronic propagators have the form of non-interacting electrons in the effective Hartree potential. In that case, the frequency sums can be performed analytically to give the Lindhard bubble. In DMFT, both the bubble $\hat{\chi}_0$ and the vertex $\hat{\Gamma}$ are matrices in fermionic frequency space and analytical treatment of the susceptibility is generically hard. However, it is known that DMFT satisfies the Ward identities and this is sufficient to derive $\mathbf{q}=0$ properties of the susceptibility~\cite{Hafermann14,Krien17}, including the Larmor frequency~\cite{KrienThesis}. 

The Ward identities express the relation between one-particle and two-particle correlation functions that has to be satisfied according to the Heisenberg equation of motion. Here, we follow the derivation given in Appendix A of Krien et al.~\cite{Krien17}, which can be generalized to systems with a Zeeman field by putting spin labels on the Green's function and dispersion in Eq.~(A6) of Krien et al.~\cite{Krien17}, and by noting that their $X$ corresponds to our $-\hat{\chi}$. Below, $\nu$ and $\omega$ are Matsubara frequencies. Note that the Zeeman field acts as a single-particle term, so $[\rho,H_\text{int}]$ still vanishes for the relevant susceptibility. Thus, the resulting Ward identity is
\begin{align}
 G^\dn_{\nu+\omega,\mathbf{k}+\mathbf{q}} - G^\up_{\nu,\mathbf{k}} = -\sum_{\mathbf{k}',\nu'} \hat{\chi}^{-+}_{\nu\nu'\omega,\mathbf{k}\mathbf{k}'\mathbf{q}} \left[ \epsilon^\dn_{\mathbf{k}'+\mathbf{q}} - \epsilon^\up_{\mathbf{k}'} -i\omega \right].
\end{align}
We get our desired result for the Larmor frequency by setting $\mathbf{q}=0$ and summing over $\mathbf{k}$ and $\nu$. On the left-hand side, the momentum and frequency sums generate expectation values, i.e., $\sum_{\mathbf{k}\nu} G^\up_{\mathbf{k},\nu} = \av{n_\up}$, while on the right-hand side $\epsilon^\dn_{\mathbf{k}'+\mathbf{q}} - \epsilon^\up_{\mathbf{k}'}=B$. Altogether,
\begin{align}
 \av{n_\dn} - \av{n_\up} &= -\chi^{-+}(i\omega,\mathbf{0}) \times \left[B-i\omega \right] \\
 \chi^{-+}(i\omega,\mathbf{0}) &= -\frac{2\av{S^z}}{i\omega-B}
\end{align}
Thus, any approximation that satisfies the Ward identities has the correct Larmor frequency. This includes both RPA and DMFT.

\subsection{Data availability}

The data that support the ﬁndings of this study are available from the corresponding author upon reasonable request.

\subsection{Code availability}
 
The python code to perform DMFT calculations for the Hubbard-Zeeman model is made available at \cite{zenodo}.

\subsection{Author contributions}

Both authors contributed equally to the work.

\subsection{Competing interests}

The authors declare no competing interests.

\bibliography{references}

\clearpage 

\begin{widetext}

\clearpage

\section{Supplementary material: Larmor precession in strongly correlated itinerant electron systems}

\renewcommand{\figurename}{Supplementary Figure}
\setcounter{figure}{0}    

\subsection{Supplementary Note 1: Homogeneous electron gas}

For the homogeneous electron gas, the Random Phase Approximation is analytically tractable. We consider electrons with kinetic energy $k^2/2m$, subject to a contact interaction $U \delta_{r,r'}$ in the magnetic channels. The correlation functions for non-interacting electrons are given by the usual Lindhard formula~\cite{GiulianiVignale}, with minor modifications due to the presence of the Zeeman field $B$ (see below): First, the energies in the spin-flip channel are shifted by $B$. Secondly, $k_{\text{F}\up}$ and $k_{\text{F}\dn}$ should be distinguished. Upon adding the contact interaction $U$ in the relevant magnetic channels, $k_{\text{F}\up}$ and $k_{\text{F}\dn}$ have to be found self-consistently, taking into account the Hartree shift of the bands due to the occupation of the opposite spin. 

The resulting susceptibility $\chi^{S^x S^x}(\omega,\mathbf{q})$ is shown in Supplementary Figure~\ref{fig:homogeneous}. It shows several of the essential features of magnetic correlation functions in the presence of a Zeeman field. There is a sharp Larmor mode with frequency $\omega_\text{L}=B$ at $\mathbf{q}=0$, as expected from the Heisenberg equation of motion. In addition, there is a electron-hole (Stoner) continuum at that goes towards the Hartree-renormalized band splitting $\omega=B_\text{eff}=B+2U\av{S^z}$ as $\mathbf{q}\rightarrow 0$, but the associated intensity necessarily vanishes in this limit. At a finite $q^\ast$, the Larmor mode meets the electron-hole continuum and becomes completely damped. Then, at $q=k_{\text{F}\up}-k_{\text{F}\dn}$, the electron-hole continuum reaches zero frequency and $\lim \chi(\omega,\mathbf{q})/\omega$ becomes finite, which is the quantity used to determine the Gilbert damping $\alpha(q)$. It remains finite up to $q=k_{\text{F}\up}+k_{\text{F}\dn}$. The fermiology behind these momentum cut-offs is illustrated in Supplementary Figure~\ref{fig:fermiology}.  

\begin{figure*}[b]
\includegraphics[width=\textwidth]{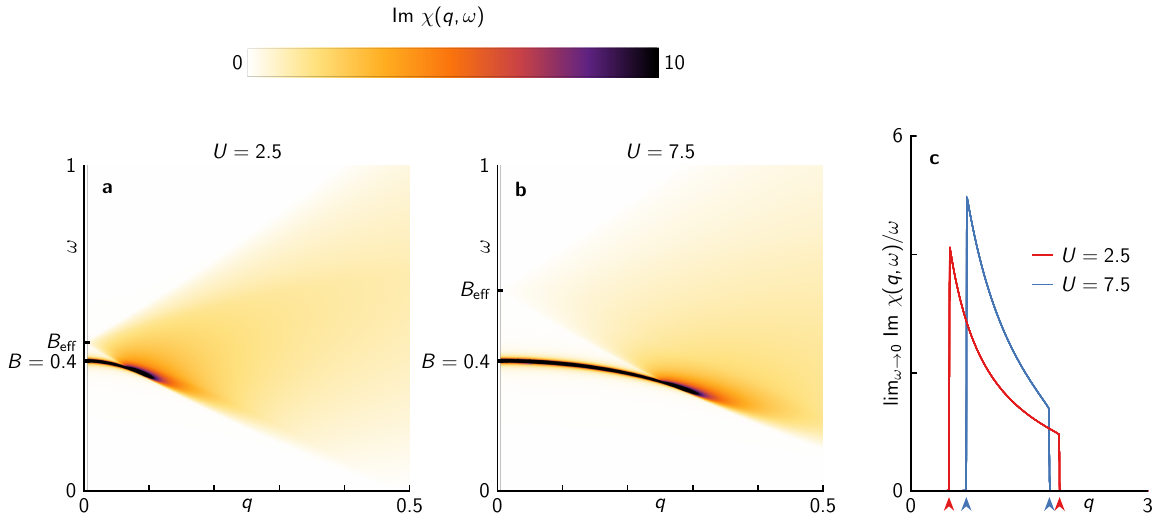} 
 \caption{Spin susceptibility of the homogeneous electron gas. The spin susceptibility $\chi^{S^x S^x}$ of the homogeneous electron gas with Zeeman field $B$, in the Random Phase Approximation. The Larmor frequency $\omega_\text{L}=B$ and $B_\text{eff}$ are indicated by a small horizontal markers. Enhancing the interaction strength makes the Larmor mode less dispersive and moves the electron-hole continuum to higher energies, resulting in a larger momentum range where the mode is visible and weakly damped. The damping is quantified in the third panel by plotting $\Im \chi(\omega,q)/\omega$ for several small values of $\omega$, $\omega \in [10^{-3},2\cdot 10^{-3}]$. All curves are on top, showing that convergence with respect to $\omega$ has been reached. Note that one should also take $\eta \ll \omega$. The damping is finite when $k_{\text{F}\up}-k_{\text{F}\dn} < q < k_{\text{F}\up}+k_{\text{F}\dn}$ (marked by arrows), where zero energy electron-hole excitations are possible. Since the $k_{\text{F}\sigma}$ are determined self-consistently, they depend on $U$, as shown by the comparison of the two curves. Furthermore, the interaction enhances the damping.}
 \label{fig:homogeneous}
\end{figure*}
 
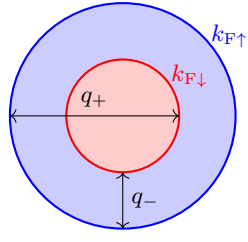
\begin{figure}
\begin{tikzpicture}
 
 \draw[thick,blue,fill=blue!20] (0,0) circle (1.5) ;
 \draw[thick,red,fill=red!20] (0,0) circle (0.75) ;
 
 \node[blue,above,right] at (45:1.5) {$k_{\text{F}\up}$};
 \node[red,above,right] at (45:0.75) {$k_{\text{F}\dn}$};
 
 \draw[<->,black] (0:0.75) -- node[above,sloped] {$q_+$} (0:-1.5) ;
 \draw[<->,black] (-90:0.75) -- node[right] {$q_-$} (-90:1.5) ;
\end{tikzpicture}
\caption{Minimal ($q_-$) and maximal ($q_+$) transferred momenta for zero-energy excitations. These determine the boundaries of the region where $\lim_{\omega\rightarrow 0} \chi^{S^x S^x}(\omega,q)/\omega \neq 0$. They are given in terms of the Fermi wavevectors $k_{\text{F}\up}$, $k_{\text{F}\dn}$}
\label{fig:fermiology}
\end{figure}

We continue with the actual calculation. The susceptibility of the homogeneous electron gas, as shown in Fig.~\ref{fig:homogeneous}, can be evaluated analytically in the usual way. Following Giuliani and Vignale (chapters 4 and 5 in Ref.\ \onlinecite{GiulianiVignale}, note that we use the sign convention $\chi=-\av{nn}$ here to be consistent with the reference), we start with
\begin{align}
 \chi_0^{-+}(\omega,\mathbf{q}) &= \int d^3 \mathbf{k} \frac{n_{\dn,\mathbf{k}+\mathbf{q}} -n_{\up,\mathbf{k}} }
 { \omega +i\eta -\epsilon_{\dn,\mathbf{k}+\mathbf{q}}+\epsilon_{\up,\mathbf{k}} }
\end{align}
Now, we split up the integral according to the numerator and change variables $\mathbf{k}\rightarrow -\mathbf{k}-\mathbf{q}$ in the first term,
\begin{align}
 \chi_0^{-+}(\omega,\mathbf{q}) &= 
 \int d^3 \mathbf{k} \frac{n_{\dn,-\mathbf{k}} }
 { \omega +i\eta -\epsilon_{\dn,-\mathbf{k}}+\epsilon_{\up,-\mathbf{k}-\mathbf{q}} } + 
 \int d^3 \mathbf{k} \frac{n_{\up,\mathbf{k}} }
 { -\omega -i\eta +\epsilon_{\dn,\mathbf{k}+\mathbf{q}}-\epsilon_{\up,\mathbf{k}} }
\end{align}
Next, we insert $\epsilon_{\dn,\mathbf{k}}=\epsilon_{k}+B/2$, $\epsilon_{\up,\mathbf{k}}=\epsilon_{\mathbf{k}}-B/2$, where $\epsilon_{\mathbf{k}}$ is the dispersion without a magnetic field.
\begin{align}
 \chi_0^{-+}(\omega,\mathbf{q}) &= 
 \int d^3 \mathbf{k} \frac{n_{\dn,-\mathbf{k}} }
 { \omega-B +i\eta -\epsilon_{-\mathbf{k}}+\epsilon_{-\mathbf{k}-\mathbf{q}} } + 
 \int d^3 \mathbf{k} \frac{n_{\up,\mathbf{k}} }
 { -\omega+B -i\eta +\epsilon_{\mathbf{k}+\mathbf{q}}-\epsilon_{\mathbf{k}} }
\end{align}
From here on, one can proceed as usual, writing $\epsilon_{\mathbf{k}+\mathbf{q}}-\epsilon_{\mathbf{k}}=q^2/2m + kq \cos(\theta)/m$, with $\theta$ the angle between $\mathbf{q}$ and $\mathbf{k}$. The factor $n_{\up,\mathbf{k}}$ picks out a spherical volume with spin-dependent radius $k_{\text{F}\sigma}$. All dimensionful constants can be pulled out and the resulting integral over $\theta$ and $\abs{k}$ leads to the special function $\Psi_d(z)=z/2+\frac{1-z^2}{4} \log \frac{z+1}{z-1}$ in three dimensions. Thus, the presence of the Zeeman field only leads to the replacement $\omega\rightarrow \omega-B$ and a different value of $k_{\text{F}\sigma}$ for the two spin flavors. The final result is
\begin{align}
 \chi_0^{-+}(\omega,\mathbf{q}) &= \frac{N_{\up}(0) k_{\text{F}\up}}{q} \Psi_d\left( \frac{\omega-B+i\eta}{k_{\text{F}\up}q}-\frac{q}{2k_{\text{F}\up}} \right) -\frac{N_{\dn}(0) k_{\text{F}\dn}}{q} \Psi_d\left( \frac{\omega-B+i\eta}{k_{\text{F}\dn}q}+\frac{q}{2k_{\text{F}\dn}} \right) \\
 \chi_0^{+-}(\omega,\mathbf{q}) &= \frac{N_{\dn}(0) k_{\text{F}\dn}}{q} \Psi_d\left( \frac{\omega+B+i\eta}{k_{\text{F}\dn}q}-\frac{q}{2k_{\text{F}\dn}} \right) -\frac{N_{\up}(0) k_{\text{F}\up}}{q} \Psi_d\left( \frac{\omega+B+i\eta}{k_{\text{F}\up}q}+\frac{q}{2k_{\text{F}\up}} \right) 
\end{align}
Here, units of $m=1$ were used, which allows us to replace $v_{\text{F}\sigma}$ by $k_{\text{F}\sigma}$. In 3d, the density of states at the Fermi level $N(0)_\sigma = \frac{k_{\text{F}\sigma}}{2\pi^2}$ in the definitions used by Giuliani and Vignale~\cite{GiulianiVignale}, where $\av{n} = \frac{1}{(2\pi)^3} \int d^3 k f(\epsilon_k)=\frac{k_F^3}{6\pi^2}$, $E=k_F^2/2$ so $N(0)=dn/dE = (dn/dk_F) (dE/dk_F)^{-1}=\frac{k_F}{2\pi^2}$.
Similarly, the total density and magnetization can be obtained from the two Fermi wavevectors as $\av{n}= \frac{1}{6\pi^2} (k_{\text{F}\up}^3 + k_{\text{F}\dn}^3)$ and $2\av{S^z}=\frac{1}{6\pi^2}(k_{\text{F}\up}^3 - k_{\text{F}\dn}^3)$.

We now consider a contact interaction in the spin channel, i.e., $U(q)=U$ independent of $\mathbf{q}$. For the RPA, we need to include the Hartree terms into $\chi^0$.  Thus, we need a self-consistent solution between the input parameters $U$, $B$ and $k_{\text{F},\up}$, $k_{\text{F},\dn}$, according to the following conditions:
\begin{align}
 \frac{1}{2} k_{\text{F}\up}^2- \frac{1}{2} k_{\text{F}\dn}^2 &= 
 B_\text{eff} = 
 B + 2 U \av{S^z} = B+ \frac{4\pi}{3} U (k^3_{\text{F},\up}-k^3_{\text{F},\dn}) \notag \\
 k^3_{\text{F},\up}+k^3_{\text{F},\dn} &= 2
\end{align}
Here, the second condition normalizes the total electron density so that $k_F=1$ at $B=0$. Alternatively, it would also be possible to fix $\mu$. Now, we get the RPA susceptibility as $\chi(\omega,\mathbf{q}) = \chi_0(\omega,\mathbf{q})/(1-U \chi_0(\omega,\mathbf{q}))$. 

\subsection{Supplementary Note 2: Spontaneously broken symmetry}

In this work, we consider paramagnetic systems in the presence of an external magnetic field. There are some similarities and difference compared to systems with spontaneously broken symmetry, such as ferromagnets and antiferromagnets, which have also been investigated using DMFT~\cite{Niyazi21}. Locally, both situations are very similar, while there are qualitative global differences. In the susceptibility, the difference between spontaneous and explicit symmetry breaking is visible at small $\mathbf{q}$. For spontaneously broken symmetries, Goldstone's theorem is essential and guarantees the presence of low-energy modes in the spin susceptibility. In the present study, there is no Goldstone mode in $\chi^{S^x S^x}$, but there are Goldstone modes present in $\chi^{S^z S^z}$ and $\chi^{nn}$ (not shown in the manuscript), or in other words in $\chi^{\up\up}$ and $\chi^{\dn\dn}$. These modes are related to the phase symmetries $c^\dagger_\sigma \rightarrow e^{i\phi} c^\dagger_{\sigma}$, which are present in the system even when $SU(2)$ symmetry is broken, and correspond to the conservation of the total particle number $\sum_a n_a$ and magnetization $\sum_a S^z_a$. In addition to the Larmor/Goldstone mode, both our system and systems with spontaneously broken symmetry have a high-energy electron-hole continuum starting at $B_\text{eff}$, which is qualitative similar in both cases. 

\subsection{Supplementary Note 3: Evolution of spectra with interaction strength}

\begin{figure*}[t]
\includegraphics[width=0.32\textwidth]{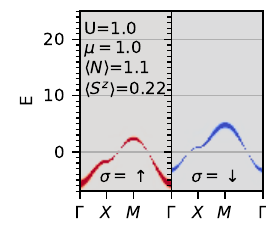}
\includegraphics[width=0.32\textwidth]{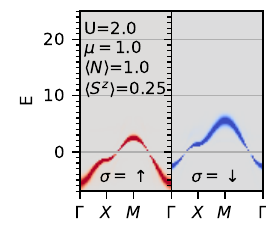}
\includegraphics[width=0.32\textwidth]{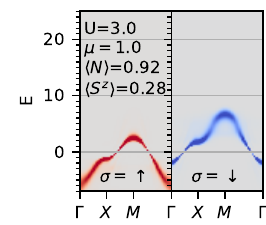} \\
\includegraphics[width=0.32\textwidth]{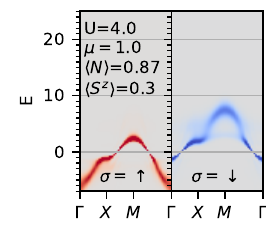}
\includegraphics[width=0.32\textwidth]{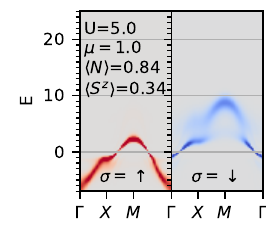} 
\includegraphics[width=0.32\textwidth]{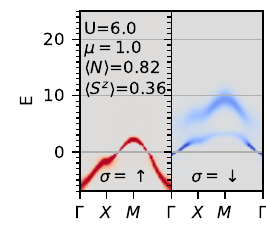} \\
\includegraphics[width=0.32\textwidth]{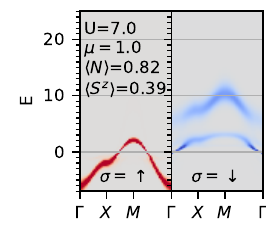}
\includegraphics[width=0.32\textwidth]{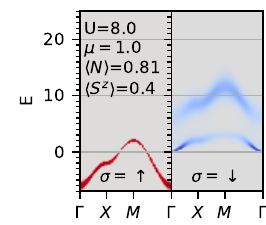}
\includegraphics[width=0.32\textwidth]{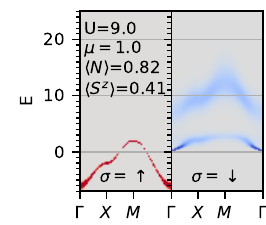} \\
\includegraphics[width=0.32\textwidth]{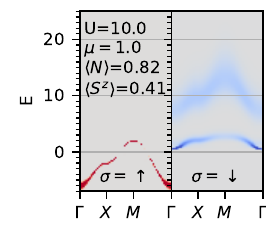}
\includegraphics[width=0.32\textwidth]{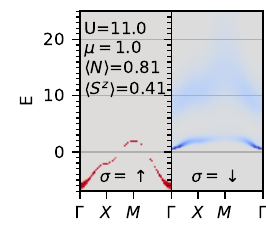}
\includegraphics[width=0.32\textwidth]{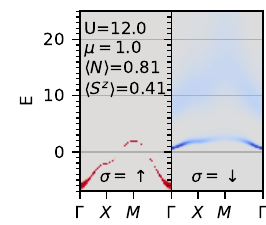}
\caption{Single-particle spectral function $A^\sigma(E,\mathbf{k})$ in DMFT, for $\mu=1$, as a function of the Hubbard interaction $U$ from $U=1$ to $U=12$. Note that the same color scales are used for all spectral functions (see Fig.~\ref{fig:susc+spec:4} in the main text), values outside this range are plotted as dark red/blue. Furthermore, very narrow bands can become invisible in the pixel map.}
\label{supp:fig:spectral2}
\end{figure*}

A larger number of single-particle spectral functions are shown in Supplementary Figure~\ref{supp:fig:spectral2}. Starting from small $U$, we find good quasiparticles for both majority ($\uparrow$) and minority ($\downarrow$) spin, which get broadened as $U$ increases. For the majority spin, the broadening then reduces once the majority band gets depopulated. At the same time, the spectral function of the minority spin channel splits in two separate dispersing branches. The higher branch is Hartree-shifted to high energy while the lower branch is pinned close to the Fermi level.

The majority and minority spin spectral functions of the fully spin polarized system are reminiscent of those found in ferromagnetic half-metals~\cite{Katsnelson08}, where the spin symmetry is broken spontaneously instead of by a magnetic field. In ferromagnetic half-metals, the high-energy minority spin band is interpreted as a normal quasiparticle with a very short lifetime due to the scattering off the majority spin electrons, while the low-energy minority states close to the Fermi level can be interpreted as non-quasiparticle states which are a superposition of a majority electron and a magnon spin wave excitation~\cite{Katsnelson08}. 

In the present model, to get some intuition for these low-energy states at large $U$, one could start with imagining frozen majority electrons in real space. In that case, if we insert a minority electron, it can hop while avoiding the sites occupied by majority electrons. The associated state is dispersive, has an energy independent of $U$, but has a reduced bandwidth, since some hopping opportunities are excluded. This picture is imperfect, since the majority electrons are not frozen and the true eigenstates upon insertion of a minority electron involves the correlated motion of the minority and majority particles, leading to scattering with momentum transfer between the spin flavors and a finite lifetime of the minority electrons. A precise investigation of the nature of the minority electron states is left for future investigations.

\clearpage

\subsection{Supplementary Note 4: Gilbert damping in real space}

The Gilbert damping $\alpha(\mathbf{q})$ can also be considered in real space, as $\alpha(\mathbf{R})$, by performing a Fourier transform of $\alpha(\mathbf{q})$. Supplementary figure~\ref{fig:alpha:r} shows the results for $U=4$ and $U=9$, i.e., moderate and strong correlations. For reference, similar results for Fe, Ni and Co are available in Figure 4 of Thonig et al.~\cite{Thonig18}. Our results show oscillations and relatively slow spatial decay. The overall magnitude of $\alpha$ as well as the spatial tail are larger in magnitude for the moderately correlated system ($U=4$), which was also visible in $\alpha(\mathbf{q})$. Note that Thonig et al.~\cite{Thonig18} considered systems with spin-orbit coupling, which leads to finite damping at $\mathbf{q}$ (as discussed in the main text) and to a much larger local than non-local $\alpha$. For our systems, $\alpha(\abs{R}=0)$ and $\alpha(\abs{R}=1)$ are more similar in magnitude, and $\sum_\mathbf{R} \alpha(\mathbf{R}) = \alpha(\mathbf{q}=0)=0$. 

\begin{figure}
 \includegraphics{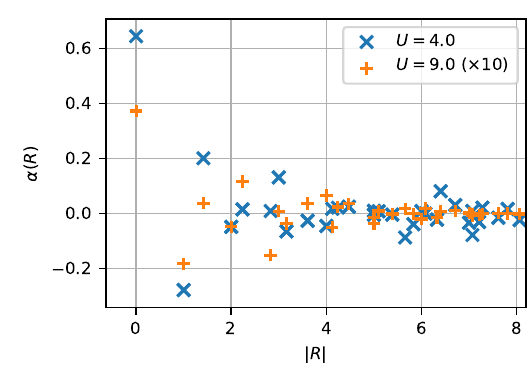}
 \caption{Gilbert damping in real space. Shown is the Gilbert damping $\alpha(R)$ as a function of the distance $\abs{R}$, for Hubbard interaction $U=4$ (blue crosses) and $U=9$ (orange plusses, enhanced by a factor $10$ for visibility). All results at chemical potential $\mu=1$. }
 \label{fig:alpha:r} 
\end{figure}

\end{widetext}

\end{document}